\documentclass[a4paper,11pt]{article}
\usepackage{pos}

\title{The four-gluon vertex in Landau gauge}

\author{Manuel Cola\c{c}o}
\author{Orlando Oliveira}
\author*{Paulo J. Silva}

\affiliation{CFisUC, Department of Physics, University of Coimbra,  3004-516 Coimbra, Portugal}

\emailAdd{manuel.sc.colaco@gmail.com}
\emailAdd{orlando@uc.pt}
\emailAdd{psilva@uc.pt}

\abstract{The Landau gauge four-gluon vertex is studied using high statistical lattice simulations for se\-ve\-ral momentum configurations.
Furthermore, the outcome of the lattice QCD simulations is compared
with calculations performed with continuum Schwinger-Dyson equations.
}

\FullConference{The 40th International Symposium on Lattice Field Theory (Lattice 2023)\\
July 31st - August 4th, 2023\\
Fermi National Accelerator Laboratory\\}

\begin{document}
\maketitle

\section{Introducion and Motivation}

In QCD, the Green functions involving only gluon fields are of utmost interest as they give important contributions, for example, to the propagators or the quark-gluon vertex --- see, for example, \cite{Huber:2018ned} for a recent review on QCD correlation functions. In perturbation theory, the Green functions with smaller number of gluon fields are well known, but their non-perturbative content is not so well studied. Indeed, in the continuum formulation of QCD and for the four gluon one-particle irreducible Green function, only a limited number of studies  using the Schwinger-Dyson equations (SDE) have been performed \cite{Kellermann:2008iw, Binosi:2014kka,Cyrol:2014kca}. Furthermore, only a single very preliminary lattice investigation is available \cite{Catumba21}. 

\begin{figure}[t]
  \begin{center}
       \includegraphics[width=5in]{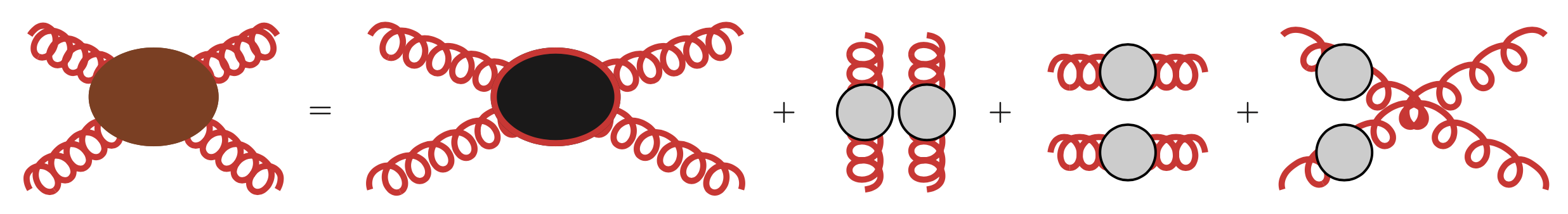} \\
       \includegraphics[width=5in]{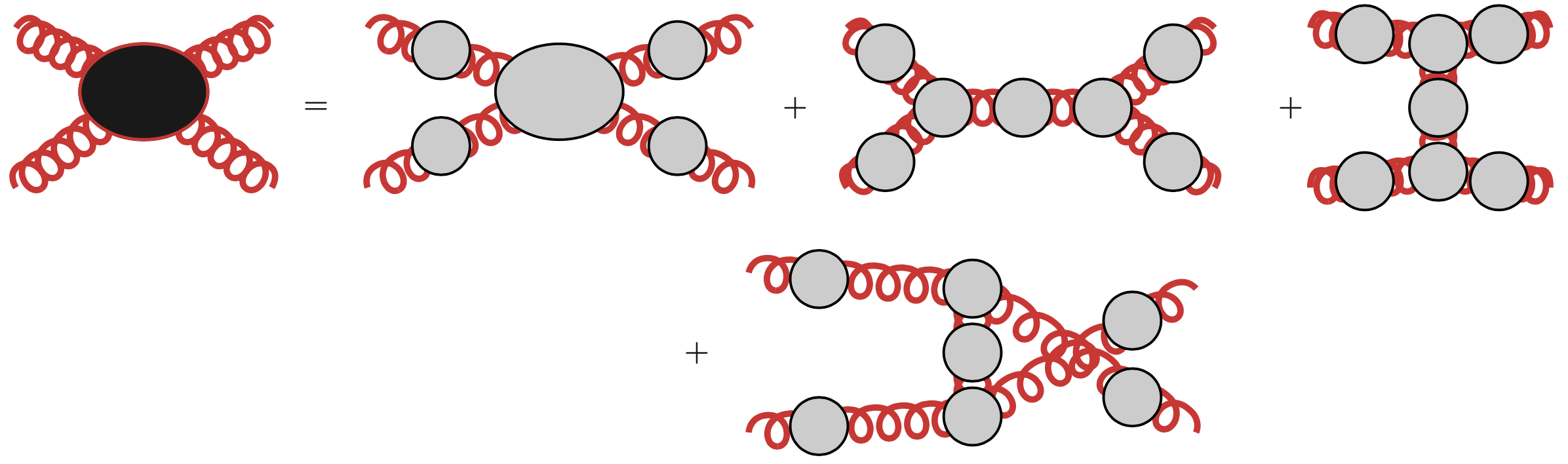} 
  \end{center}
 \caption{Four gluon Green function: the full (upper plot) and the connected (lower plot) functions written in terms of one-particle irreducible Green functions.}
\label{Fig:GreenF}
\end{figure}

The color-Lorentz structure of the four-gluon one-particle irreducible Green function is rather complex and its full description, for all possible kinematics, requires more than one hundred tensors to assemble a complete basis. Any computation of this vertex using continuum approaches becomes quite involved and simplifications have to be introduced. On the other hand, its computation with lattice simulations is challenging not only because this Green function has a larger number of external legs, demanding the use of very large ensembles to have reliable output, but also because on the lattice only the full Green functions are accessed. In Fig. \ref{Fig:GreenF} we detail the four gluon Green functions. Indeed, disentangling the one-particle irreducible form factors from the full Green function is another challenge. However the extraction of the one-particle irreducible Green functions has been done for e.g. the three gluon, the ghost-gluon and the quark-gluon Green functions (see, for example, \cite{Pinto-Gomez:2022brg,Duarte:2016ieu,Catumba:2021hng,Kizilersu:2021jen,Cucchieri:2008qm,Sternbeck:2006rd} and references there in) and can also be also done, with a proper choice of kinematics, for the four point function.

\section{The amputated Green functions from lattice simulations}

On the lattice, to access the one-particle irreducible function one aims to suppress the contribution of the disconnected diagrams. For the four-gluon function,
this can be done by a right choice of kinematics. Indeed, as can be seen in Fig. \ref{Fig:GreenF}, the condition $p_i + p_j \ne 0$ on the external momenta
avoids the disconnected diagrams but one still has to remove the contributions coming from diagrams with three gluon irreducible terms.  
In the Landau gauge, due to the orthogonality of the gluon propagator, setting all the momenta proportional removes the diagrams
that include three-gluon vertices; see the lower plot of the Fig. \ref{Fig:GreenF}. Then, our choice of momenta being
\begin{equation}
\begin{array}{l@{\hspace{1cm}}l@{\hspace{1cm}}l@{\hspace{1cm}}l@{\hspace{4cm}}l}
  p_1 = 0, & p_2 = p, & p_3 = p , & p_4 = -2 \, p, \\
  p_1 = 0, & p_2 = p, & p_3 = 2 \, p , & p_4 = -3 \, p, \\
  p_1 = p, & p_2 = p, & p_3 = p , & p_4 = -3 \, p,
\end{array}
\nonumber 
\end{equation}
where $p$ is a generic momentum.
For these kinematics, the full four-point Green function can be decomposed as
\begin{small}
\begin{eqnarray}
& &
~~ \hspace*{-1cm} \mathcal{G}^{abcd}_{\mu\nu\eta\zeta} (p_1, \, p_2, \, p_3, \, p_4)  =  
\Big( P_\perp (p_1) \Big)_{\mu\mu^\prime} \, \Big( P_\perp (p_2) \Big)_{\nu\nu^\prime} \, 
\Big( P_\perp (p_3) \Big)_{\eta\eta^\prime} \, \Big( P_\perp (p_4) \Big)_{\zeta\zeta^\prime}  
 ~ D(p^2_1) \, D(p^2_2) \, D(p^2_3) \, D(p^2_4) \nonumber \\
& & \qquad\qquad\qquad
\Bigg( F(p^2_1, \dots) \,t^{(0)}  \,  ^{abcd}_{\mu^\prime\nu^\prime\eta^\prime\zeta^\prime}    ~ + ~  
          G(p^2_1, \dots) \,t^{(1)} \,  ^{abcd}_{\mu^\prime\nu^\prime\eta^\prime\zeta^\prime}    ~ + ~  
          H(p^2_1, \dots) \,t^{(2)} \,  ^{abcd}_{\mu^\prime\nu^\prime\eta^\prime\zeta^\prime}   ~ + ~ \cdots \Bigg)
           \nonumber 
\end{eqnarray}
\end{small}
where the Landau gauge propagator is
\begin{eqnarray}
D^{ab}_{\mu\nu} (p)  ~ = ~   \delta^{ab}\Big( P_\perp (p) \Big)_{\mu\nu} \,  \,  D(p^2) 
 ~ = ~  \delta^{ab} \left( \delta_{\mu\nu} - \frac{p_\mu p_\nu}{p^2}  \right)\,  \,  D(p^2) 
\end{eqnarray}
and $t^{(i)}$ are the tensor basis for the one-particle irreducible four gluon Green function. Of the possible tensors that define the complete
basis, in the following, we will consider only  three different tensor structures $t^{(i)}=\widetilde{\Gamma}^{(i)}\,,\, i=0,1,2$, where the $\widetilde{\Gamma}$ are given by
\begin{eqnarray}
\widetilde{\Gamma}^{(0)} \,  ^{abcd}_{\mu\nu\eta\zeta}  & = &
   f_{abr} f_{cdr} \left( \delta_{\mu\eta} \delta_{\nu \zeta} - \delta_{\mu\zeta} \delta_{\nu \eta} \right) ~ + ~
       f_{acr} f_{bdr} \left( \delta_{\mu\nu} \delta_{\eta\zeta} - \delta_{\mu\zeta} \delta_{\nu \eta}   \right)  \nonumber \\
       & & \qquad\qquad + ~
       f_{adr} f_{bcr} \left( \delta_{\mu\nu} \delta_{\eta\zeta}  - \delta_{\mu\eta} \delta_{\nu \zeta}  \right) \ ,
            \\
\widetilde{\Gamma}^{(1)} \,  ^{abcd}_{\mu\nu\eta\zeta}  & = &
   d_{abr} d_{cdr} \left( \delta_{\mu\eta} \delta_{\nu \zeta} + \delta_{\mu\zeta} \delta_{\nu \eta} \right)  ~ + ~
       d_{acr} d_{bdr} \left( \delta_{\mu\nu} \delta_{\eta\zeta} + \delta_{\mu\zeta} \delta_{\nu \eta}   \right)  \nonumber \\
       & & \qquad\qquad + ~
       d_{adr} d_{bcr} \left( \delta_{\mu\nu} \delta_{\eta\zeta}  + \delta_{\mu\eta} \delta_{\nu \zeta}  \right)  \ ,
  \\
\widetilde{\Gamma}^{(2)} \,  ^{abcd}_{\mu\nu\eta\zeta}   & = &
\Big(  \delta^{ab}  \, \delta^{cd} + \delta^{ac}  \, \delta^{bd} + \delta^{ad}  \, \delta^{bc}  \Big) 
~
\Big(  \delta_{\mu\nu}  \, \delta_{\eta\zeta} + \delta_{\mu\eta}  \, \delta_{\nu\zeta} + \delta_{\mu\zeta}  \, \delta_{\nu\eta} 
\Big) \ ,
\end{eqnarray}
where the first is associated with tree level Feynman rule, the second is obtained from the first replacing $f_{abc} \rightarrow d_{abc}$ and
symmetrising the operator and the third is a tensor that completes the basis to describe the $p, \, p, \, p, \, - 3p$ kinematics as discussed in
\cite{Binosi:2014kka}. The form factors measured in the simulation are
\begin{eqnarray}
F^{(i)} (p^2) &  = &
 \widetilde{\Gamma}^{(i)} \,  ^{abcd}_{\mu\nu\eta\zeta}   ~~ \mathcal{G}^{abcd}_{\mu\nu\eta\zeta} (p_1, \, p_2, \, p_3, \, p_4)
 \ . 
 \nonumber
\end{eqnarray}

\section{The lattice setup and the lattice form factors}

The gauge configurations used in the calculation are sampled with the Wilson action. After sampling, each configuration is
rotated to the Landau gauge by maximising, over the gauge orbits, the functional
\begin{equation} 
   F(U^{g}) =   \mathrm{Re}  \mathrm{Tr} \sum_{x, \mu}\left[
   \; g(x) \, U_{\mu}(x) \, g^{\dagger}(x+ \hat{\mu}) \right] \ .
\end{equation}
At a maximum of $F$ the condition $\partial_{\mu} A^{a}_{\mu} = 0 + \mathcal{O}(a^2)$ is fulfilled and, once more, up to corrections of order $a^2$ the
lattice definitions match the continuum. 

The results of the simulations discussed below use 4620 configurations defined on a $32^4$ lattice at $\beta = 6.0$, that has 
a lattice spacing of $a = 0.1016(25)$ fm measured from the string tension.
The lattice simulations use Chroma \cite{Edwards} and PFFT \cite{Pippig}.

\begin{figure}[t]
  \begin{center}
    \vspace*{-2mm}
\includegraphics[width=0.62\textwidth]{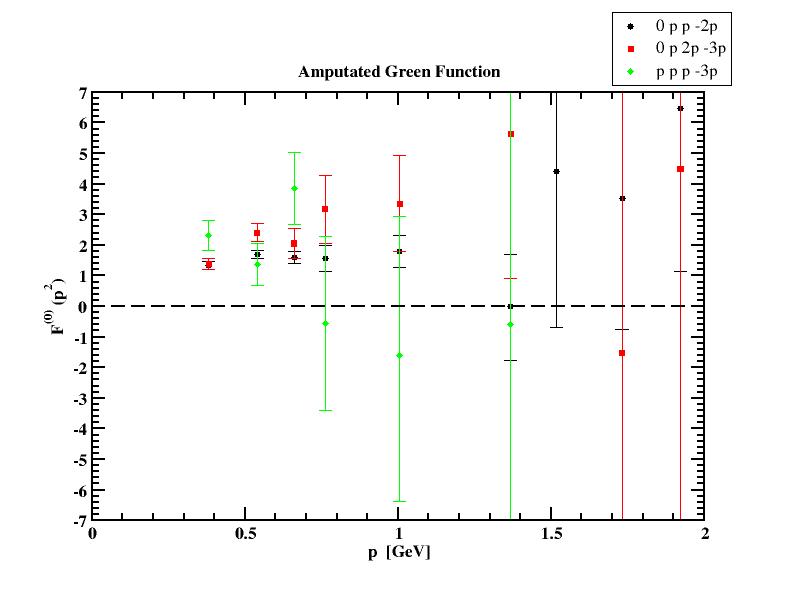} \\
\includegraphics[width=0.62\textwidth]{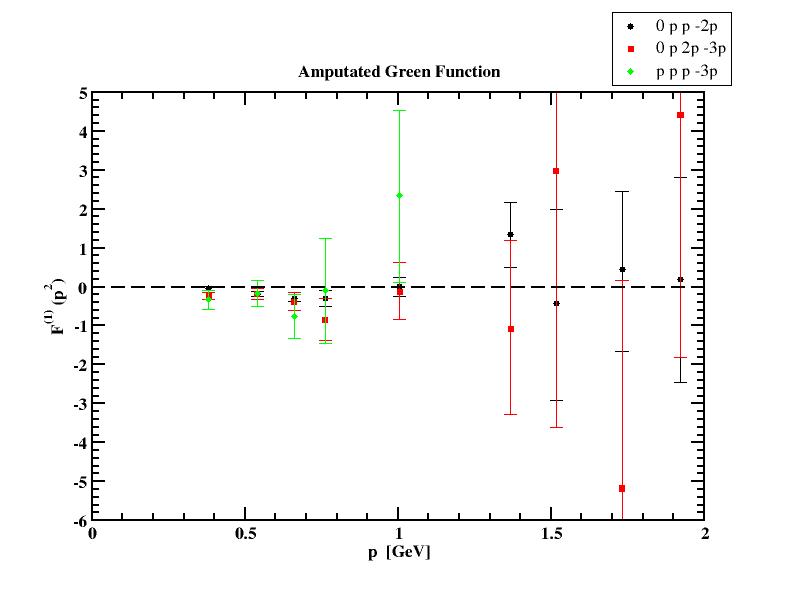} \\
\includegraphics[width=0.62\textwidth]{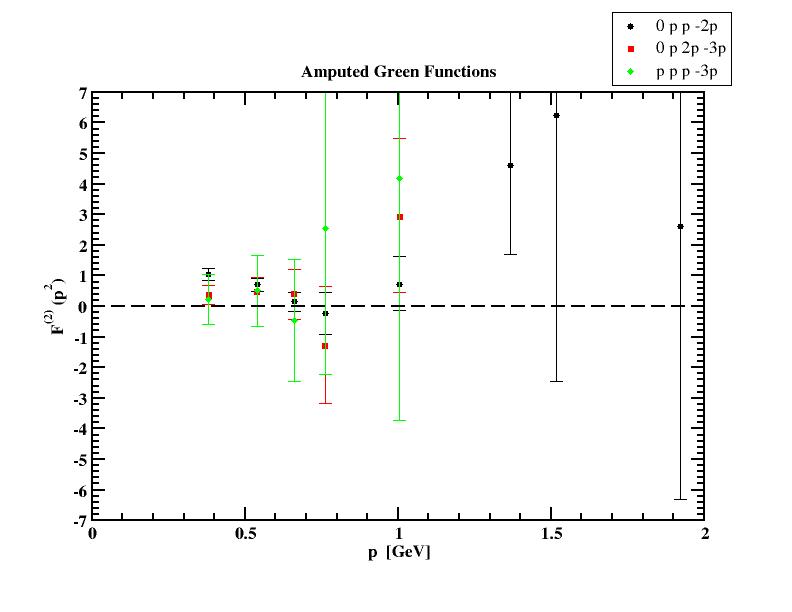}
\end{center}
\caption{The bare lattice amputated four gluon Green functions.}
\label{LattFig}
\end{figure}

For the $32^4$ lattice, the bare amputated four gluon Green functions are reported in Fig. \ref{LattFig}. We call the reader's attention that 
for comparison between lattice and continuum results a global constant factor should be included for all the form factors.
As Fig. \ref{LattFig} shows, the tree level structure seems to be associated with the dominant form factor and, for the range of momenta accessed in the
simulation, it seems that the four gluon one particle irreducible vertex is constant. The lattice form factors shows no clear sign of IR divergence, a result
that should be read with care as the minimum momentum accessed in the simulation is $\sim$ 380 MeV.

\section{Continuum Schwinger-Dyson results for the four gluon vertex}

\begin{figure}[t]
\begin{center}
\includegraphics[width=0.55\textwidth]{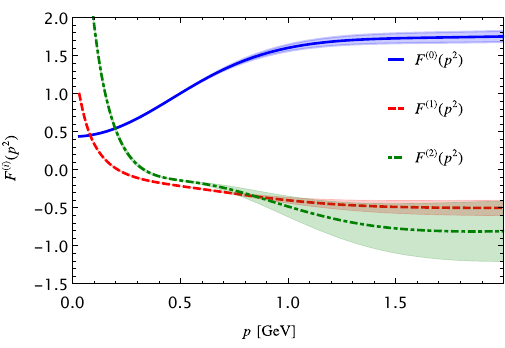} 
\end{center}
\caption{Form factors from the solutions of the Schwinger-Dyson equations.}
\label{Fi-SDE}
\end{figure}

For comparison with the lattice data, our colleagues  A. C. Aguilar, M. N. Ferreira, J. Papavassi\-liou and L. R. Santos provided us with the
results of a Schwinger-Dyson equation (SDE) calculation using a one-loop dressed approximation for the same form factors
$F^{(i)}(p^2)$. The outcome of the SDE calculation can be seen in Fig. \ref{Fi-SDE}.

The results of the SDE calculation show that all of the $F^{(i)}$ are nearly constant for momenta above $p\gtrapprox 1$~GeV, and are suppressed
 for $p\sim 500$~MeV. Moreover,  $F^{(0)}(p^2)$ is finite at $p = 0$ and the form factors $F^{(1)}(p^2)$ and $F^{(2)}(p^2)$ exhibit an infrared
 logarithmic infrared divergence that is driven by the ghost-loops. 
 The logarithmic divergences are only apparent for $p<300$~MeV, that is below the range of momenta accessible in the lattice simulation.

\section{Conclusions and Outlook}

The lattice simulation discussed here should be read as preliminary and, therefore, the comparison with the  continuum calculation is only possible 
at a qualitative level. As shown, the lattice data and the results of solving the SDE are, at a qualitative level, in good agreement 
for all the  form factors. Although not shown here, the lattice data for all the form factors is compatible with 
``planar degeneracy"  \cite{Pinto-Gomez:2022brg} that assumes a dependence only on  $s^2 = \sum_i p^2_i$ instead of $p^2$.

The outcome of this exploratory work is encouraging and we aim to explore the same kinematics with larger ensembles of configurations as well as
further kinematics, in the near future, and to enlarge the set of form factors to be computed.

\section*{Acknowledgements}

\noindent
The authors acknowledge Arlene C. Aguilar, Mauricio N. Ferreira, Joannis Papavassiliou and Leonardo R. Santos for helpful discussions.
The authors acknowledge the computing time provided by the Laboratory for Advanced Computing at the University of Coimbra 
(FCT contracts  2021.09759.CPCA and  2022.15892.CPCA.A2). Work  supported by FCT contracts UIDB/04564/\-2020, UIDP/04564/2020 and CERN/FIS-PAR/0023/2021. P.J.S.  acknowledges
financial support from FCT  under Contract
No. CEECIND/00488/2017.

\clearpage

\begin{thebibliography}{99}


\bibitem{Huber:2018ned}
M.~Q.~Huber,
Phys. Rept. \textbf{879} (2020), 1.


\bibitem{Kellermann:2008iw}
C.~Kellermann and C.~S.~Fischer,
Phys. Rev. D \textbf{78} (2008), 025015.


\bibitem{Binosi:2014kka}
D.~Binosi, D.~Iba\~nez and J.~Papavassiliou,
JHEP \textbf{09}, 059 (2014).

\bibitem{Cyrol:2014kca}
A.~K.~Cyrol, M.~Q.~Huber and L.~von Smekal,
Eur. Phys. J. C \textbf{75}, 102 (2015).

\bibitem{Catumba21}
G.~T.~R.~Catumba,
arXiv:2101.06074 [hep-lat].



\bibitem{Pinto-Gomez:2022brg}
F.~Pinto-G\'omez.
F.~De Soto, M.~N.~Ferreira, J.~Papavassiliou and J.~Rodr\'\i{}guez-Quintero,
Phys. Lett. B \textbf{838}, 137737 (2023).

\bibitem{Catumba:2021hng}
G.~T.~R.~Catumba, O.~Oliveira and P.~J.~Silva,
PoS \textbf{LATTICE2021} (2022), 467.

\bibitem{Duarte:2016ieu}
A.~G.~Duarte, O.~Oliveira and P.~J.~Silva,
Phys. Rev. D \textbf{94} (2016) no.7, 074502.

\bibitem{Kizilersu:2021jen}
A.~K\i{}z\i{}lers\"u, O.~Oliveira, P.~J.~Silva, J.~I.~Skullerud and A.~Sternbeck,
Phys. Rev. D \textbf{103} (2021) no.11, 114515.

\bibitem{Cucchieri:2008qm}
A.~Cucchieri, A.~Maas and T.~Mendes,
Phys. Rev. D \textbf{77} (2008), 094510.

\bibitem{Sternbeck:2006rd}
A.~Sternbeck,
arXiv:hep-lat/0609016 [hep-lat].




\bibitem{Edwards}
R. G. Edwards \textit{et al.},
Nucl. Phys. B Proc. Suppl. 140, 832 (2005). 

\bibitem{Pippig}
M. Pippig, SIAM J. Sci. Comput. 35, C213 (2013).


\end{thebibliography}
\end{document}